%% file: main.tex
\documentclass[conference]{IEEEtran}
\usepackage{cite}
\usepackage{amsmath,amssymb,amsfonts}
\usepackage{algorithmic}
\usepackage{graphicx}
\usepackage{textcomp}
\usepackage{xcolor}

\usepackage[caption=false, font=footnotesize]{subfig}

\def\BibTeX{{\rm B\kern-.05em{\sc i\kern-.025em b}\kern-.08em
    T\kern-.1667em\lower.7ex\hbox{E}\kern-.125emX}}

\usepackage[utf8]{inputenc}
\usepackage{booktabs}
\usepackage{multirow}
\usepackage{tabularx}
\usepackage{tabulary}
\usepackage{url}
\usepackage{soul}
\usepackage[np, autolanguage]{numprint}
\usepackage{verbatim}

\usepackage{tikz}
\newcommand*\circled[1]{\tikz[baseline=(char.base)]{
            \node[shape=circle,draw,inner sep=1pt,font=\sffamily\footnotesize] (char) {#1};}}
\usepackage{pifont}

\usepackage{pdfcomment}
\usepackage{float}

\usepackage[all=normal,floats,leading, paragraphs,charwidths,bibbreaks,mathspacing,wordspacing]{savetrees}

\newcommand{\lmttfont}{\fontfamily{lmtt}\selectfont}

\begin{document}

\title{Enhancing Failure Propagation Analysis\\in Cloud Computing Systems}




\author{
\IEEEauthorblockN{Domenico Cotroneo, Luigi De Simone, Pietro Liguori, Roberto Natella}
\IEEEauthorblockA{\textit{Università degli Studi di Napoli Federico II} \\  
Naples, Italy\\
\{cotroneo, luigi.desimone, pietro.liguori, roberto.natella\}@unina.it}
\and
\IEEEauthorblockN{Nematollah Bidokhti}
\IEEEauthorblockA{\textit{Futurewei Technologies, Inc.} \\
USA \\ 
nbidokht@futurewei.com}
}


\maketitle
\thispagestyle{plain}
\pagestyle{plain}

\begin{abstract}
In order to plan for failure recovery, the designers of cloud systems need to understand how their system can potentially fail. Unfortunately, analyzing the failure behavior of such systems can be very difficult and time-consuming, due to the large volume of events, non-determinism, and reuse of third-party components. To address these issues, we propose a novel approach that joins fault injection with anomaly detection to identify the symptoms of failures. 
We evaluated the proposed approach in the context of the OpenStack cloud computing platform. We show that our model can significantly improve the accuracy of failure analysis in terms of false positives and negatives, with a low computational cost.
\end{abstract}




\section{Introduction}
\label{sec:introduction}

\input{intro.tex}

\section{Problem Statement}
\label{sec:problem}
\input{problem_statement.tex}

\section{Proposed Methodology}
\label{sec:method}
\input{proposed_methodology.tex}

\section{Experimental Evaluation}
\label{sec:evaluation}
\input{experimental_evaluation.tex}

\section{Related Work}
\label{sec:related}
\input{related.tex}

\section{Conclusion}
\label{sec:conclusion}
\input{conclusion.tex}

\section*{Acknowledgments}
This work has been partially supported by the PRIN 2015 project ``GAUSS'' funded by MIUR (Grant n. 2015KWREMX\_002) and by UniNA and Compagnia di San Paolo in the frame of Programme STAR.

\bibliographystyle{IEEEtran}
\bibliography{bibliography}

\end{document}

%% file: intro.tex

Cloud computing is becoming an attractive solution for running services with high-reliability requirements, such as in the telecom and healthcare domains \cite{sultan2014making,kuo2011opportunities,doukas2012bringing,yin2015joint}. 
However, cloud computing systems are often exposed to unpredictable failure conditions \cite{8497007}. These failures can propagate across several components or layers of the system (e.g., storage, virtual network, compute instances, etc.) in complex ways, leading to cascading effects (\emph{failure propagation}) that make recovery actions more problematic.

Therefore, identifying and analyzing failure propagation is an important activity to design more effective recovery actions. 
Fault injection is a relevant approach, which emulates faults to anticipate worst-case scenarios, such as network partitions, high network latency, replica crashes, and I/O exceptions \cite{Ju2013a,Pham2011,pham2017failure,Gunawi2011a,Joshi2011b,cerveira2015recovery}. Fault injection has reached a level of maturity that it is routinely used to reveal failures in real-world  systems, including cloud computing software such as key-value data stores and distributed computing frameworks (e.g., Cassandra, ZooKeeper) \cite{Gunawi2011a}, entire cloud computing services (e.g., streaming services deployed by Netflix) \cite{chaos_monkey} and infrastructures (e.g., IaaS providers such as Amazon) \cite{limoncelli2012resilience}. 

Nevertheless, there are still open issues for its adoption in cloud systems. 
Indeed, as the scale and the complexity of these systems increase, it becomes harder for developers to identify (and to analyze) failures that are triggered by fault injection. Furthermore, failure propagation analysis too often relies on the knowledge, the experience, and the intuition of human analysts since existing fault injection solutions provide limited support to the analyst for understanding what happened during an experiment \cite{natella2016assessing}. 

The current state of practice is to detect failures (e.g., service unavailability, performance degradation) by monitoring the quality of service during the fault injection test; more sophisticated solutions detect failures by monitoring properties expressed with formal specifications, such as finite state machines \cite{deligiannis2016uncovering}, relational logic \cite{Gunawi2011a}, and special-purpose languages \cite{reynolds2006pip}. 
However, once a service failure has been triggered by fault injection and detected by monitoring mechanisms, a human analyst still needs to analyze the chain of events (e.g., messages) that occurred among the location where the fault/error is injected and the component that experiences the service failure. Yet, this failure analysis still relies on intuition and manual effort of the human analyst \cite{cotroneo2018enhancing}. Unfortunately, manual analysis is too difficult and time-consuming, because of:

\begin{itemize}
    \item The \textbf{high volume of messages} generated by large distributed systems that the human analyst needs to scrutinize;
    
    \item The \textbf{non-determinism in distributed systems}, in which the timing and the order of messages can unpredictably change even if there is no failure, which introduces noise in the analysis, and increases the effort of the human analyst to pinpoint the failure (i.e., to discriminate the anomalies caused by a fault from genuine variations of the system);
    
    \item The use of \textbf{``off-the-shelf'' software components}, either proprietary or open-source (such as application frameworks, middleware, data stores, etc.), whose events and protocols can be difficult to understand and to manually analyze.
\end{itemize}

This work aims to provide automated support for analyzing failures triggered by fault injection in cloud computing systems. We aim to avoid the human analyst to manually inspect thousands of events, by automatically identifying the few relevant events that are related to the injected fault, while discarding noisy, uninteresting events. To this goal, we propose an approach that extends fault injection, by combining it with black-box tracing and anomaly detection for failure analysis. The driving idea is to train a \emph{probabilistic model} of the events in the distributed system under test under \emph{fault-free} conditions, by using variable-order Markov Models for analyzing event sequences. Afterward, the system is tested with fault injection, and event traces are collected under these \emph{faulty} conditions. The faulty event traces are analyzed with anomaly detection by using the probabilistic model, and the anomalous events are reported to the human analyst for understanding how to avoid failures. 

We experimentally evaluate the proposed approach in the context of the OpenStack cloud management platform, which is the basis for many commercial cloud management products \cite{OpenStackProducts}, and it is widespread both among public cloud infrastructure providers and private users \cite{OpenStackUsers}. Our experiments show that the proposed approach can be applied to event traces that are generated by a large, ``off-the-shelf'' distributed system, without relying on  knowledge of its internals, with a low rate of false positives (i.e., genuine variations are not mistaken for failure symptoms) and of false negatives (i.e., actual anomalies caused by a fault are not missed), and with a low computational cost.

In the following of this paper, Section~\ref{sec:problem} elaborates on the problem addressed by this paper, and provides a motivating example; Section~\ref{sec:method} presents the proposed methodology for failure analysis; Section~\ref{sec:evaluation} experimentally evaluates the methodology; Section~\ref{sec:related} discusses related work; Section~\ref{sec:conclusion} concludes the paper.

%% file: problem_statement.tex
To better understand the research problem addressed by this paper, we discuss an example of fault-injection experiment on the OpenStack cloud computing platform.
OpenStack is a cloud management system that controls large pools of computing, storage, and networking resources in a data center. It provides a dashboard and APIs that can be used both by cloud operators to manage the infrastructure, and by end-users to offer resources as-a-service. 

In our fault-injection experiments, we inject faults into the three most important services of OpenStack \cite{denton2015learning,solberg2017openstack}: (i) the \textbf{Nova} subsystem, which provides services for provisioning instances (VMs) and handling their life cycle; (ii) the \textbf{Cinder} subsystem, which provides services for managing block storage for virtual instances; and (iii) the \textbf{Neutron} subsystem, which provides services for provisioning virtual networks, including resources such as \emph{floating IPs}, \emph{ports} and \emph{subnets} for instances. In turn, these subsystem includes several components (e.g., the Nova sub-system includes \emph{nova-api}, \emph{nova-compute}, etc.), which interact through message queues internally to OpenStack. The Nova, Cinder, and Neutron sub-systems provide external REST API interfaces to cloud users.

\begin{figure}[t!]
\centering
\includegraphics[width=1.0\columnwidth]{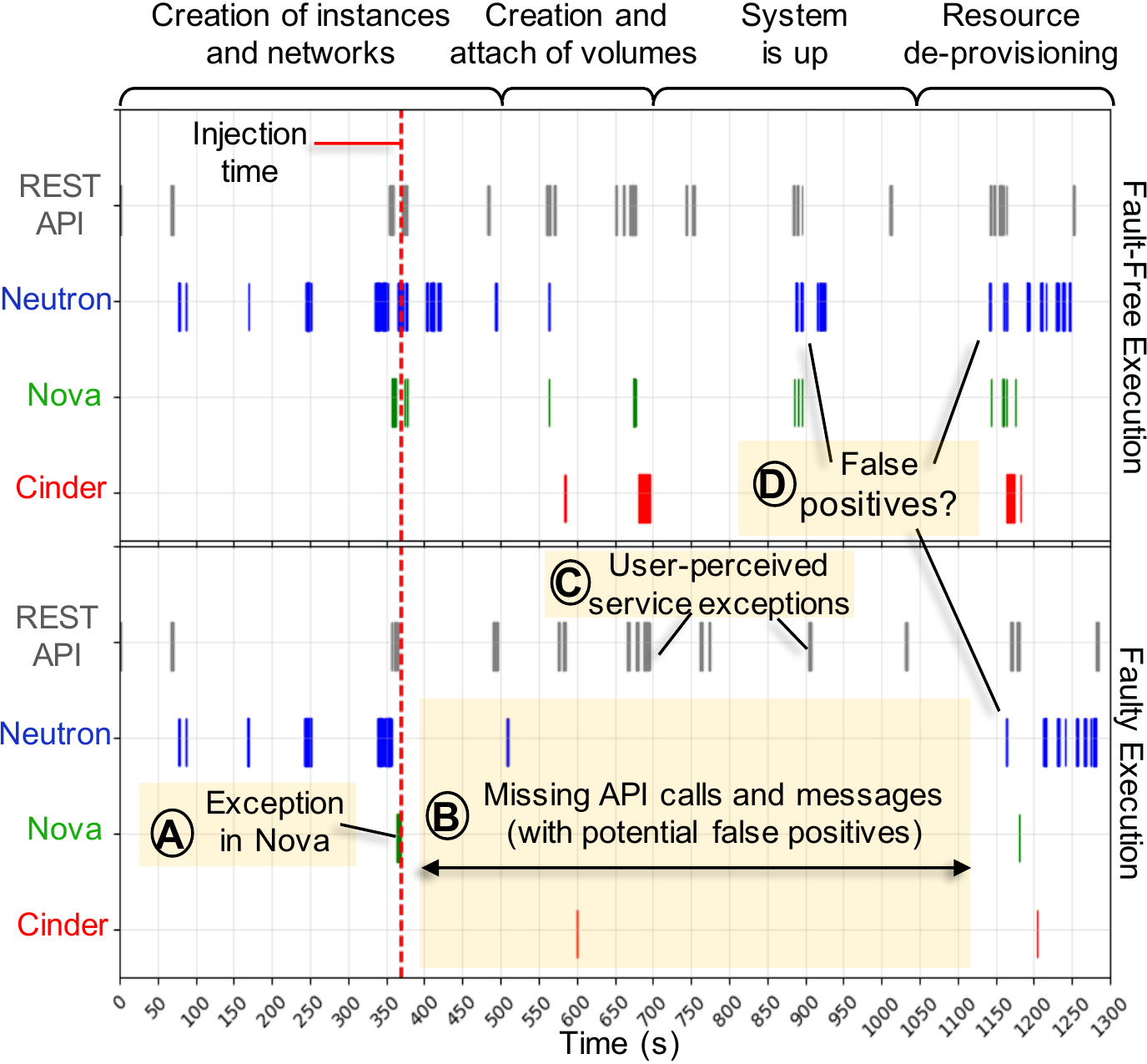}
\vspace{-0.7cm}
\caption{A graphical representation of a fault-injection experiment.}
\label{fig:Example_Failure}
\vspace{-0.8cm}
\end{figure}

A simple graphical representation of a fault-injection experiment is shown in \figurename{}~\ref{fig:Example_Failure}.
This representation shows remote procedure calls that are made for communication in the distributed system. These calls are displayed as intervals over the timeline of the experiment. We consider both API calls between the client and the OpenStack REST APIs (the topmost sequence of calls), and internal API calls within OpenStack, which are performed by Nova, Neutron, and Cinder using message queues (the other three sequences of calls). 
In order to see the effects of the injected fault, we show two subplots: the former shows a normal execution of the system (\emph{fault-free execution}), in which no fault is injected; the latter shows the execution of the system when a fault is injected in the Nova subsystem (\emph{faulty execution}). 
Since both executions are performed under the same conditions (i.e., same software and hardware configuration, same workload, etc.), any deviation between the faulty and the fault-free execution is considered an anomaly due to the injected fault.


The workload used in this example first creates several resources (i..e, networks, instances, volumes, etc.), then it performs basic operations in order to stimulate the different components of the system (e.g., attaching a volume to an instance, check the connectivity, reboot an instance, etc.) before cleaning up the created resources.  All these operations are performed by invoking the OpenStack APIs. 

One of these API calls is an asynchronous request for creating a new VM instance. After the API call ends, OpenStack Nova takes a few minutes for creating and initializing the instance. During these operations, we inject a Python exception in order to force a failure (\circled{A}).

\figurename{}~\ref{fig:Example_Failure} points out that there are several API calls in the fault-free execution that are missing in the faulty execution (\circled{B}) since the injected fault causes a failure that affects several OpenStack subsystems over a relatively long time period. Indeed, Nova does not complete the initialization of the VM instance due to the fault, leaving the VM in an inactive state. Moreover, the OpenStack Neutron subsystem was also unable to attach the virtual network to the VM instance. Later on (i.e., after about five minutes) the workload client experienced a service exception when calling the API of the Cinder subsystem, which manages storage volumes in OpenStack (\circled{C}). Consequently, the workload could not attach the volume to the VM instance. Both Nova and Neutron do not raise any API exception, but the failure only became apparent to the client when invoking the API of the Cinder subsystem. Therefore, the issue propagates both across subsystems (from Nova to Neutron and Cinder) and across time, since the client perceives the failure only after a relatively long time. This behavior is problematic from the point of view of high-availability, and thus of defining proper recovery actions, as the propagation delay also increases the time-to-detect and the time-to-recover the failure. Furthermore, the longer the propagation chain the more difficult will be for a developer reasoning about how to best tolerate the fault, e.g., whether to manage the fault in Nova, Neutron and/or Cinder and at which time to manage the fault during the workload. For example, the API could return a more timely notification of the failure to the client, either by introducing a callback mechanism in the Nova API that creates the instance or by returning an error from other API calls to Nova or Neutron.

\begin{figure*}[!ht]
\centering
\includegraphics[width=1.9\columnwidth]{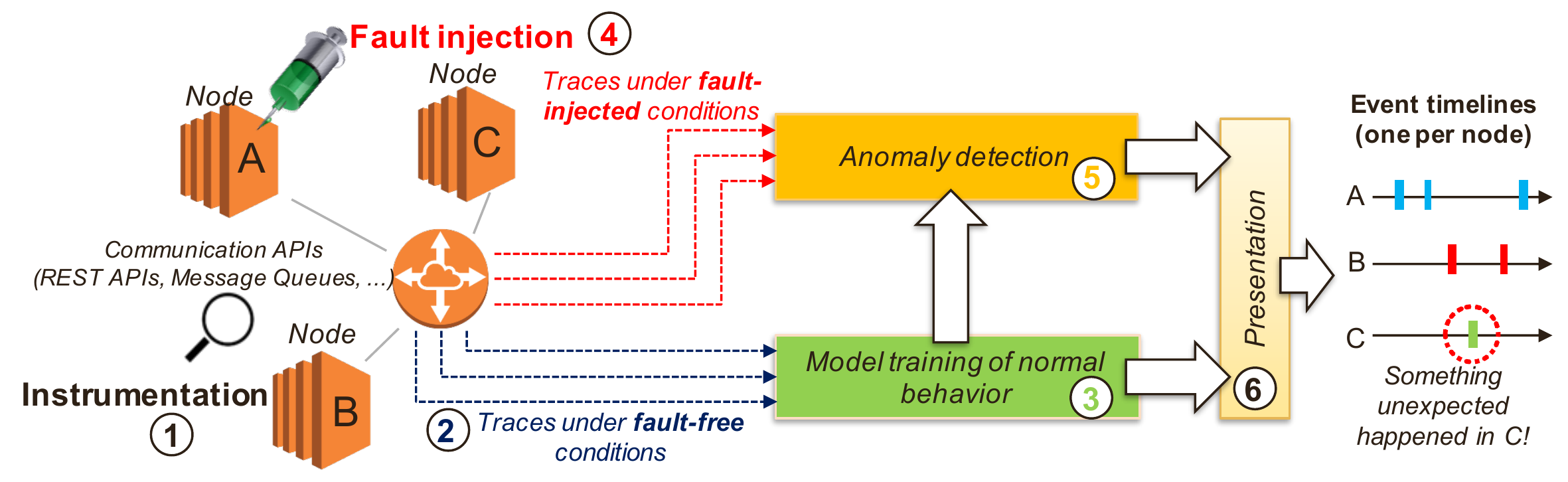}
\vspace{-0.4cm}
\caption{Overview of the proposed approach.}
\label{fig:approach_overview}
\vspace{-0.6cm}
\end{figure*}

The analysis of a fault-injection experiment can be inaccurate due to the non-determinism of the API calls in distributed systems.
For example, the Neutron subsystem uses asynchronous messages and polling for distributing state updates across its components, thus such messages could be easily misclassified as anomalies. Moreover, due to the asynchronous nature of several APIs, it is difficult to properly identify whether API calls order does not matter (i.e., is due to non-determinism) or should be carefully taken into account because of the failure. In this point, \figurename{}~\ref{fig:Example_Failure} also highlights events that could be false positives (\circled{D}), both among the fault-free and the faulty execution. Thus, we need to understand if the differences among such two executions are due to the non-determinism in the system (i.e., they are not related to the failure) or not (i.e., they are actually anomalies). Considering the false positives makes the debugging more difficult and cumbersome for the human analyst, as each execution may include hundreds of API calls to analyze with only a few ones relevant for understanding the failure.

In this work, we propose an algorithm for enhancing the failure propagation analysis of the fault-injection experiments, by adopting a rigorous probabilistic approach to pinpoint unlikely messages that are related to the failure, with the goal of achieving high accuracy in identifying the true anomalies.

%% file: proposed_methodology.tex
\begin{figure*}[!htb]
\centering
\includegraphics[width=2\columnwidth]{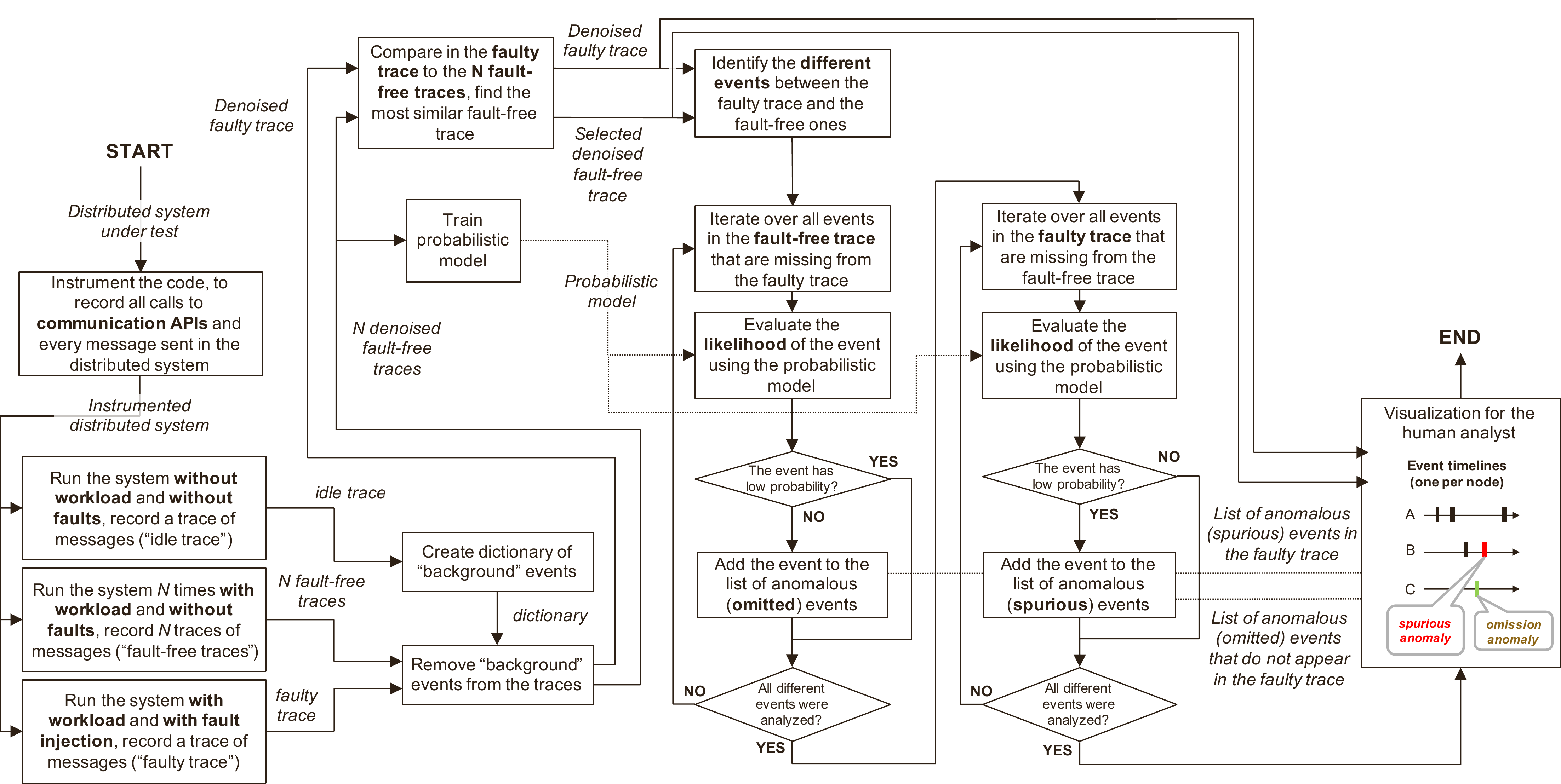}
\vspace{-0.3cm}
\caption{Detailed workflow of the proposed approach.}
\label{fig:approach_flowchart}
\vspace{-0.6cm}
\end{figure*}


\figurename{}~\ref{fig:approach_overview} shows an overview of the approach.
Firstly, we instrument the communication APIs of the system (step \circled{1}). We consider a distributed system as a set of black-box components that interact with each other via public service interfaces (e.g., REST APIs, message queues).
Therefore, we exercise the system by applying a workload without injecting any fault (step \circled{2}). We record all messages exchanged among the components, and between the components and the workload client. These messages constitute the \emph{fault-free trace}. Several fault-free traces are collected by executing the same workload several times, to take into account the natural variability of such traces.

In order to have an accurate model of fault-free system behavior, we define a probabilistic model that is trained by the fault-free traces (step \circled{3}). Due to non-determinism, this model considers the ``benign'' variability of the interactions (e.g., different ordering, type, or duration of events) that can occur under fault-free conditions. After training the model, a fault injection experiment is performed in the distributed system (step \circled{4}), for each fault encompassed in the analysis. This step will produce the so-called \emph{fault-injected traces} (also \emph{faulty traces}), i.e., one per experiment. 
The faulty traces are then analyzed by the proposed probabilistic model in order to detect the actual deviation(s), i.e., the anomaly (ies),  from the normal behavior (step \circled{5}). 

In order to emphasize messages that were omitted because of the injected fault (i.e. only occurring in fault-free conditions), and new messages that were caused by the injected fault (i.e., only occurring under faulty conditions), the results of anomaly detection  are visualized by presenting to the human analyst the messages of both the fault-injected and of a fault-free execution (step \circled{6}).

Figure \ref{fig:approach_flowchart} shows a detailed flowchart of the proposed approach. In the following of this section, we discuss the individual steps of the workflow.

\subsection{Instrumentation and pre-processing}


The proposed approach records and analyzes \emph{messages} that occur in the distributed system. 
Messages are the key observation point for debugging and verification of distributed systems, as they reflect well the activity of the distributed system \cite{leesatapornwongsa2016taxdc}. For example, nodes perform work when they receive message requests (e.g., through remote procedure calls), and reply with messages for providing responses and results; moreover, nodes use messages to asynchronously notify a new state to other nodes in the distributed system.

Thus, the first step of our approach consists in instrumenting the distributed system under test, in order to keep track of the messages that are sent between nodes during a test. 
In general, it is possible to get traces of the communication among components leveraging \emph{run-time tracing techniques}, which allow us to instrument the source- or binary-code and record the execution of specific points in the software. 
In particular, our approach gets information about messages by collecting traces of \emph{communication API invocations} made by the distributed software. For example, in our approach we instrument the calls to APIs of popular middleware technologies such as \emph{REST frameworks} (e.g., Django \cite{django} and Spring \cite{spring}) and \emph{message queueing} (e.g., AMQP \cite{ampq} and RabbitMQ \cite{rabbitmq}).
Example of tracing toolkits are Zipkin \cite{zipkin} (used in this paper), Jaeger \cite{jaeger}, and Appdash \cite{appdash}. These monitoring tools are familiar to developers of distributed systems, as they are already used for debugging, performance monitoring and optimization, root cause analysis, and service dependency analysis \cite{chow2014mystery, chen2002pinpoint}. 


This instrumentation is a form of ``black-box tracing'' since it does not require any knowledge about the internals of the system under test, but only which are the communication APIs used by the system. This approach is suitable when testers do not have a full and detailed understanding of the entire distributed system; this is the case of distributed systems developed by large teams (in which testers and developers might be distinct people), and distributed systems that embed components developed by third-parties.


The approach records the beginning and the end of every call to the communication APIs by inserting a probe using the distributed tracing system. The tracer records all information about the exchanged messages, such as the time at which the communication API has been called and its duration, the component that invoked the API (\emph{message sender}), and the remote service that has been requested through the API call (\emph{called service}).
We refer to the calls to communication APIs as \emph{events}; thus, the execution of the distributed system generates an \emph{event trace}.
The approach orders the events in the trace with respect to the timestamp of the event collector. Our anomaly detection technique is designed to be tolerant to the non-determinism (e.g., due to random messaging delays) of the events by using a probabilistic technique, which will be discussed in the section \ref{subsec:probabilistic_modeling}.


This lightweight approach for event collection allows us to deploy tracing with low intrusiveness and does not require more detailed information about the internals of the system. For example, the tracer within the components does not need to collect and propagate a session identifier across messages related to the same session, which would require the human analyst to customize the data collection according to the specific application \cite{osprofiler} or to collect more extensive information at the OS- and network-level \cite{gu2018kerep, barham2003magpie}.




\subsection{Data collection}

Once the distributed system has been instrumented, it is executed several times to perform fault injection tests. 
The distributed system is monitored during test execution; at the same time, the system is stimulated with a workload (e.g., by generating client requests), and a fault is injected into the system. Each test injects a different fault, and only one fault is injected per test. For each test, we collect a message trace (\textbf{fault-injected trace}).

In addition to fault-injected traces, we also execute the system and collect traces without fault injection (\textbf{fault-free traces}). In general, collecting fault-free traces (also known as \emph{golden runs} or \emph{reference runs}) is a common practice in fault injection experiments since they are used as a reference to understand how the system derailed from a correct execution due to the injected fault \cite{hsueh1997fault,leeke2009evaluating,natella2016assessing}. While previous studies used this approach for the analysis of non-distributed systems (such as embedded systems), we extend this approach to distributed systems, by addressing the problems of non-determinism and scalability of the analysis. 
To collect fault-free traces, the approach executes the system $N$ times, by running the same workload used in fault injection tests, but without injecting any fault. The messages exchanged in each execution are stored in a fault-free trace, i.e., one fault-free trace per workload execution. These $N$ fault-free traces are then used for training the model of ``normal'' behavior of the distributed system. The model will be used as a reference for analyzing failures. 
We use more than one fault-free trace since the model needs to reflect the variability of the execution that characterizes distributed systems (e.g., the relative ordering of messages). We expect that the larger is the number of training traces, the more accurate is the model that represents the normal behavior. The use of the fault-free traces is discussed more in detail in the next subsections, and the impact of the number of training traces is empirically evaluated in Section~\ref{sec:evaluation}.

Finally, during data collection, we need to take into account the messages that are not due to the workload but they are independently generated by background and asynchronous activities in the system at arbitrary times. For example, these messages represent events that are internally produced by garbage collection, resource monitoring, updating database indexes, etc.. Since these messages are not strictly related to the workload, they can (mistakenly) appear as anomalies during fault injection tests. 
For this reason, our approach properly removes such unrelated events. Therefore, we perform a preliminary analysis of the system in which no workload is applied. The approach keeps the system idle for a few minutes before and after a fault-free execution of the workload and records any background message into a trace (\textbf{idle trace}). Then, we use the idle trace to create a dictionary of \emph{background events} that will be ignored in the subsequent analysis of the fault-free and fault-injected traces. 

\subsection{Trace comparison}
\label{subsec:trace_comparison}

Internally, the approach represents the events within a trace with unique identifiers (i.e., \emph{symbols}), so that two events of the same type are identified by the same symbol.
In particular, we assign a unique symbol to every distinct pair $<$\emph{message sender}, \emph{called service}$>$ (e.g., $<$\emph{Cinder}, \emph{attach volume}$>$). 
Thus, the event traces are converted into \emph{sequences of symbols}. 

In order to identify the differences between the faulty and the normal execution of the system, the approach performs a string comparison on the fault-injected sequence and one of the fault-free sequences. In particular, the approach looks for the \emph{longest common subsequence} (LCS) of such sequences \cite{bergroth2000survey}. 
The LCS is a subset of symbols that are present in both sequences in the same order, and that can be obtained by removing (a minimal number of) symbols from the original sequences. This kind of problem is recurrent in computer science, such as in bioinformatics and in source code versioning (e.g., in the \emph{diff} Unix tool), and can be solved with efficient algorithms \cite{hunt1976algorithm,myers1986nd}.

To perform the comparison, the approach selects one fault-free trace among the ones collected at the beginning of the workflow (\figurename{}~\ref{fig:approach_flowchart}). In particular, the approach selects the fault-free trace \emph{most similar} to the fault-injected trace since we want to identify and to filter out from the failure analysis as much \emph{common} events as possible (i.e., the approach aims to discard the subset of messages that also happen with the same type and order in at least one fault-free sequence), in order to focus the attention of the human analyst on the \emph{anomalous} events (i.e., the differences between the faulty and the most similar fault-free sequence). 

The similarity between two strings $x$ and $y$ is measured by considering the length of the LCS (\(|LCS(x,y)|\)) \cite{budalakoti2009anomaly}, i.e., the number of symbols that appear in both strings while preserving the order of symbols. In particular, we compute the normalized length of the LCS \((nLCS)\), where
%
$nLCS(x,y)=\frac{|LCS(x,y)|}{\sqrt{l_x \cdot l_y}}$ 
and where $l_x$ and $l_y$ are the lengths of the individual strings $x$ and $y$. The approach uses this metric to identify the fault-free trace of the training set most similar to the fault-injected trace (\emph{selected fault-free trace}).

\subsection{Probabilistic modeling}
\label{subsec:probabilistic_modeling}

The analysis performed with LCS is still prone to inaccuracies since there may be differences between the fault-injected trace and the selected fault-free trace that are caused by non-deterministic reorderings, and thus are not related to failures. These differences lead to \emph{false positives} that may divert the attention of the human analyst. 
To overcome this problem, the approach uses a \emph{Markov model} to estimate the probability of an event, in order to evaluate whether the event is anomalous in a probabilistic sense. Markov modeling is a popular approach for the probabilistic analysis of sequences of symbols (e.g., to predict the probability of a future symbol), such as in bioinformatics \cite{stanke2003gene}, data compression \cite{rissanen1983universal}, and text and speech recognition \cite{rabiner1989tutorial}. In our context, we evaluate the probability of the events marked as anomalies in the previous comparison performed with LCS. Thus, we use the probabilistic model as a further reference to analyze the anomalous events. Such a model takes into account the ``benign'' variations in the ordering and type of messages that happen in fault-free conditions.

We opt for a Markov model where the states are a direct representation of the observed events. However, a simple Markov chain still does not suffice for our purposes, since the probability of the next state (i.e., the next event of the sequence) would only depend on the current state (i.e., the \emph{memoryless} property). In general, this is not the case for a sequence of events that can be generated by a distributed system; in practice, the probability of an event is correlated with the history of the previous events. For example, in the case of the OpenStack platform, the occurrence of an event representing a ``volume attach'' operation must be preceded by a sequence of several preliminary operations on the volume and on the instance to be attached (e.g., an instance must be created and initialized before attaching volume).

Ultimately, we decide to use \emph{higher-order} Markov models, where the probability of events takes into account the history of the previous states of a sequence.
In particular, since conditioning random variables could vary based on the specific observed realization, we adopt \emph{Variable-order Markov Models} (VMMs). VMMs estimate the probability that a symbol \(\sigma\) can appear after a sequence $s$ (named \emph{context}), by counting the joint occurrences of \(\sigma\) and $s$ in the training sequence to build the predictor \(\hat{P}\), for variable cardinalities of $s$ \cite{begleiter2004prediction}.



In this work, we use the notation defined by Begleiter et al. \cite{begleiter2004prediction}.
Let \(\Sigma\) be a finite alphabet. A learner is given a training sequence \(x_1^n = x_1 x_2 ... x_n\), where \(x_i \in \Sigma\) and \(x_i x_{i+1}\) is the concatenation of \(x_i\) and \(x_{i+1}\). Based on \(x_1^n\), the goal is to learn a model \(\hat{P}\) that provides a probability assignment for any future outcome given the past.
Specifically, for any context \(s \in \Sigma\) and symbol \(\sigma \in \Sigma\), the learner should generate a conditional probability \(\hat{P}(\sigma|s)\). 
The accuracy of the predictor \(\hat{P}(\cdot|\cdot)\) is typically measured by its average log-loss \(l(\hat{P},x_i^T)\) with respect to a test sequence \(x_1^T = x_1 ... x_T\):
\vspace{-0.2cm}
\begin{equation}
    \ell(\hat{P},x_i^T)= -\frac{1}{T} \sum_{i=1}^{T} \log \hat{P}(x_i | x_1...x_{i-1})
\end{equation}
\vspace{-0.2cm}




There exist many algorithms in the scientific literature for training and applying VMMs \cite{begleiter2004prediction}. 
Our approach uses the \emph{Prediction by Partial Matching - Method C} (PPM-C) lossless compression algorithm \cite{cleary1997unbounded}, which is a variant of the original PPM algorithm published in 1984 by Cleary and Witten \cite{cleary1984data} that includes a set of improvements proposed by Moffat \cite{moffat1990implementing}. 
PPM is a finite-context statistical modeling technique that builds a predictor by combining several fixed-order context models \cite{cleary1997unbounded}, with different values of the order, ranging from zero to an upper bound $D$ (i.e., the maximal order of the Markov model) \cite{mavadati2014comparing}.
For more detailed information on PPM and the Method C variant, we refer the reader to the work of Begleiter et al. \cite{begleiter2004prediction}.

In this work, we set the maximum order of the VMMs, i.e. $D$, by measuring the number of events that can be triggered by an individual request from a client of the distributed system. In response to a client request, the distributed system generates a sequence of messages among its internal components, until it reaches again a quiescent state, or it returns a reply to the client. Since an event is most likely influenced by the previous events in the context of the same client request, we set $D$ to the maximum number of events triggered by a client request. This choice is conservative since this number (e.g., several tens of events in our case study) tends to be much higher than the context length chosen in previous studies on VMMs \cite{cleary1997unbounded}. 



\subsection{Classification procedure}

The ultimate result of the proposed approach is to classify the events into:

\begin{itemize}
    \item\textbf{Common events}: Events that occurred both in the fault-injected trace and in \textit{at least one} of the fault-free traces, with the same type and order.
    
    \item \textbf{Anomalous events}: Differences between the fault-injected trace and all of the fault-free traces. These events are further classified into:
    \begin{itemize}
        \item \textbf{Spurious events}: Events that would normally not happen under fault-free conditions.
        \item \textbf{Missing events}: Events that happen in fault-free conditions, but do not happen under fault injection.
    \end{itemize}
    
\end{itemize}

The approach trains the VMM by using a set of $n-1$ fault-free traces (i.e., all the fault-free traces, except the \emph{selected fault-free trace} with the highest similarity to the fault-injected trace). Then, we apply the VMM to compute the probabilities of events, in order to determine whether they are anomalous. Specifically, the approach performs two steps:

\vspace{3pt}
\noindent
$\rhd$ \emph{Analysis of LCS differences that only appear in the fault-injected trace.} 
In the first step, the fault-injected trace takes the role of the \emph{test sequence} for the VMM. We focus on symbols of the test sequence that were highlighted as differences in the previous LCS analysis. The goal is to confirm whether these symbols are actually unlikely events, not only with respect to the selected fault-free trace (i.e., the one used for determining the LCS) but also according to the whole set of fault-free traces in the training set. For each event not included in the LCS, we compute the probability of the event according to the VMM. If the probability is lower than a threshold \(\epsilon_{\textrm{SPURIOUS}}\), then the symbol has a low likelihood to appear in that position of the sequence; thus, the VMM confirms that the symbol represents a \textbf{spurious} anomalous event. Otherwise, the event is considered non-anomalous.

\vspace{3pt}
\noindent
$\rhd$ \emph{Analysis of LCS differences that only appear in the selected fault-free trace.} 
In the second step, the \emph{selected fault-free trace} takes the role of the \emph{test sequence} for the VMM. As for the previous step, we focus on symbols of the test sequence that were highlighted as differences in the previous LCS analysis. In this case, we consider the events that only appear in the selected fault-free trace: therefore, from the point of view of the fault-injected trace, these events represent \emph{omissions}. This step confirms whether these omissions are indeed likely, and thus should be considered anomalies. The approach applies the VMM to the events that only appear in the fault-free trace, by computing the probabilities of such events according to remaining fault-free traces in the dataset. If the probability of the event is higher than a threshold \(\epsilon_{\textrm{MISSING}}\), then there is a high likelihood for the symbol to be in that position of the sequence. Therefore, the fact that the event is missing in the fault-injected trace should be considered an anomaly, and thus it is marked as a \textbf{missing} anomalous event. Otherwise, if the probability of the event is not high, then the lack of the event from the fault-injected trace is considered non-anomalous.

\vspace{3pt}

We remark that even if the two steps perform similar comparisons, the results obtained by them are different and complementary. If the fault-injected trace contains an anomalous event with a \emph{low probability value} according to the VMM, then it is confirmed as spurious. Similarly, if the fault-injected trace does not contain an event with a \emph{high probability value} in the selected fault-free trace, then the event is confirmed to be an omission. 
A practical approach is to select conservative thresholds (e.g., \(\epsilon_{\textrm{SPURIOUS}} = 20\%\) and \(\epsilon_{\textrm{MISSING}} = 80\%\)), so that the VMM can filter out most of the LCS differences that are not actually spurious/missing events; and to leave to the human analyst the decision about the uncertain events. Therefore, the accuracy of the probabilistic model is an important factor that makes this approach suitable in practice. The accuracy of our approach is further analyzed in the rest of the paper.

%% file: experimental_evaluation.tex
In this section, we evaluate the approach in the context of fault injection experiments in the OpenStack platform. In \S{}~\ref{subsec:setup}, we present the experimental setup, and in \S{}~\ref{subsec:accuracy} and \S{}~\ref{subsec:performance} we report on the accuracy and performance of the proposed approach.

\subsection{Experimental Setup}
\label{subsec:setup}


In our fault-injection experiments, we targeted OpenStack version 3.12.1 (release \emph{Pike}), deployed on Intel Xeon servers (E5-2630L v3 @ 1.80GHz) with 16 GB RAM, 150 GB of disk storage, and Linux CentOS v7.0, connected through a Gigabit Ethernet LAN.

We injected faults during the execution of OpenStack components, by simulating exceptional conditions during the interactions between components. We targeted the internal APIs used by OpenStack components for managing instances, volumes, networks, and other resources. For example, we injected faults during calls to the \emph{nova-compute} component within the Nova subsystem to manage new instances. The injected faults represent exceptional cases, e.g., a resource that is not found or unavailable, a processing delay when retrieving a resource, or an incorrect value caused by the user, the configuration, or a bug inside OpenStack. In particular, we considered the following kind of faults:

\begin{itemize}

    \item \textbf{Throw exception}: An exception is raised on a method call, according to pre-defined, per-API list of exceptions;
    
    \item \textbf{Wrong return value}: A method returns an incorrect value. In particular, the returned value is corrupted according to its data type (e.g., we replace an object reference with a null reference, or replace an integer value with a negative one);
    
    \item \textbf{Wrong parameter value}: A method is called with an incorrect input parameter. Input parameters are corrupted according to the data type, as for the previous fault type;

    \item \textbf{Delay}: A method is blocked for a long time before returning a result to the caller. This fault can trigger timeout mechanisms inside OpenStack or can cause a stall.
    
\end{itemize}

 \vspace{2pt}
We performed three distinct fault injection campaigns, in which we applied three different workloads described in the following.

    
    \vspace{1pt}
    \noindent
    $\rhd$ \textbf{New deployment workload} (DEPL): This workload configures a new virtual infrastructure from scratch, by stimulating all of the target subsystems (i.e., Nova, Neutron, and Cinder) in a balanced way. 
    This workload creates VM instances, along with key pairs and a security group; attaches the instances to an existing volume; creates a virtual network consisting in a subnet and a virtual router; assigns a floating IP to connect the instances to the virtual network; reboots the instances, and then deletes them;
        
    \vspace{1pt}
    \noindent
    $\rhd$ \textbf{Network management workload} (NET): 
    This workload includes network management operations, in order to stress more the Neutron subsystem and virtual networking. The workload initially creates a network and a VM, then generates network traffic via the public network. After that, it creates a new network with no gateway, brings up a new network interface within the instance, and generates traffic to check whether the interface is reachable. Finally, it performs a router rescheduling, by removing and adding a virtual router resource;
    
    \vspace{1pt}
    \noindent
    $\rhd$ \textbf{Storage management workload} (STO): This workload performs storage management operations on instances and volumes, in order to stress more the Nova and Cinder subsystems. In particular, the workload creates a new volume from an image, boots an instance, then rebuilds the instance with a new image (e.g., as it would happen for an update of the image). Finally, it performs a cleanup of the resources. 
    
 \vspace{2pt}

All of these workloads invoke the OpenStack APIs, which are provided by the Nova, Cinder, and Neutron subsystems. We implemented the workloads by reusing integration test cases from the \emph{OpenStack Tempest} project \cite{openstack_tempest}, since these tests are already designed to trigger several subsystems and components of OpenStack and their virtual resources. We selected this kind of workload in order to point out propagation effects across subsystems that may be caused by fault injection.

In-between calls to service APIs, our workload generator performs \emph{assertion checks} on the status of the virtual resources, in order to reveal failures of the cloud management system. 
In particular, these checks assess the connectivity of the instances through SSH and query the OpenStack API to ensure that the status of the instances, volumes, and the network is consistent with the expectation of the tests. In the context of our methodology, assertion checks serve as \emph{ground truth} about the occurrence of failures during the experiments (i.e., a reference for evaluating the accuracy of the proposed approach). 
We consider an experiment as failed if at least one API call returns an error (\textbf{API error}) or if there is at least one assertion check failure (\textbf{assertion check failure}). 
Before every experiment, we clean-up any potential residual effect from the previous experiment, in order to ensure that the potential failure is only due to the current injected fault. To this end, we re-deploy the cloud management system, remove all temporary files and processes, and restore the OpenStack database to its initial state.


In order to find all the injectable locations in Nova, Neutron, and Cinder, we performed a full scan of the source code according to the fault types described above. Then, for each workload, we identified the injectable locations that were covered by the workloads (i.e., we run the workload without injecting anything), and we performed one fault injection test per covered location. 
In total, we performed $2,137$ fault injection tests, and we observed failures in $1,432$ tests (67\%).
In the remaining tests (33\%), there were neither API errors nor assertion failures, since the fault did not affect the behavior of the system (e.g., the corrupted state is not used in the rest of the experiment). This is a typical phenomenon that often occurs in fault injection experiments \cite{christmansson1996generation,lanzaro2014empirical}; yet, the experiments provided us a large and diverse set of failures for our analysis.

Table~\ref{tab:dictionary_wl} shows, for each workload, the number of \emph{unique events} (i.e., the events with different pair $<$\emph{message sender}, \emph{called service}$>$) observed in the distributed system during the execution of the workloads, the average length of the fault-free sequences (in term of number of events in the trace), the total number of fault injection experiments for the workload, and the number of experiments that experienced at least one failure.
The number of unique events and the total number of events reflects the extent and diversity of the work put on the system. We notice that DEPL is the most extensive workload in terms of both distinct operations and the total number of operations, followed by NET and by STO. 
These differences among the workloads are meant to evaluate the approach under different levels of complexity and non-determinism.


\begin{table}[t]
\caption{Workload characteristics}
\label{tab:dictionary_wl}
\centering
\begin{tabular}{>{\centering\arraybackslash}p{1cm} >{\centering\arraybackslash}p{1cm} >{\centering\arraybackslash}p{2cm} >{\centering\arraybackslash}p{1cm} >{\centering\arraybackslash}p{1cm} >{\centering}p{3cm}}

\toprule

\textbf{Workload} & \textbf{Num. unique events} & \textbf{Avg. num. of events per fault-free trace} & \textbf{Num. of total exps.} &  \textbf{Num. of failed exps.} \\ 
\midrule

DEPL & 53 & 243 & 945 & 449\\ 
NET & 37 & 212 & 290 & 206\\ 
STO & 34 & 85 & 902 &  777\\ 

\bottomrule

\end{tabular}
\vspace{-0.6cm}
\end{table}

We used the distributed tracer \emph{Zipkin} for collecting message traces. We instrumented the following communication points:

\begin{itemize}
    
    \item The \emph{OSLO Messaging library}, which uses a message queue library, by exchanging messages with an intermediary queuing server (RabbitMQ) through RPC messages. These messages are used for communication among OpenStack subsystems;
    
    \item The \emph{RESTful API libraries} of each OpenStack subsystem, i.e., the \emph{novaclient} for Nova (it implements the OpenStack Compute API \cite{computeAPI}), the \emph{neutronclient} for Neutron (it implements the OpenStack Network API \cite{networkAPI}), and the \emph{cinderclient} for Cinder (it implements the OpenStack Block Storage API \cite{storageAPI}). These interfaces are used for communication between OpenStack and its clients.
    
\end{itemize}

In total, we instrumented only $5$ selected functions of these components (e.g., the {\lmttfont cast} method of OSLO to broadcast messages), by adding very simple annotations only at the beginning of these methods, for a total of 20 lines of code. We neither added any further instrumentation to the subsystems under test nor used any knowledge about OpenStack internals.

\subsection{Accuracy evaluation}
\label{subsec:accuracy}

\begin{figure*}[!htb]

    \centering
    \includegraphics[keepaspectratio=true, width=0.2\textwidth]{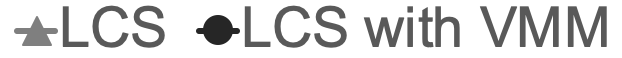}

    \subfloat[New deployment workload (DEPL).\label{fig:fp_generic_wl}]{%
        \includegraphics[width=0.67\columnwidth]{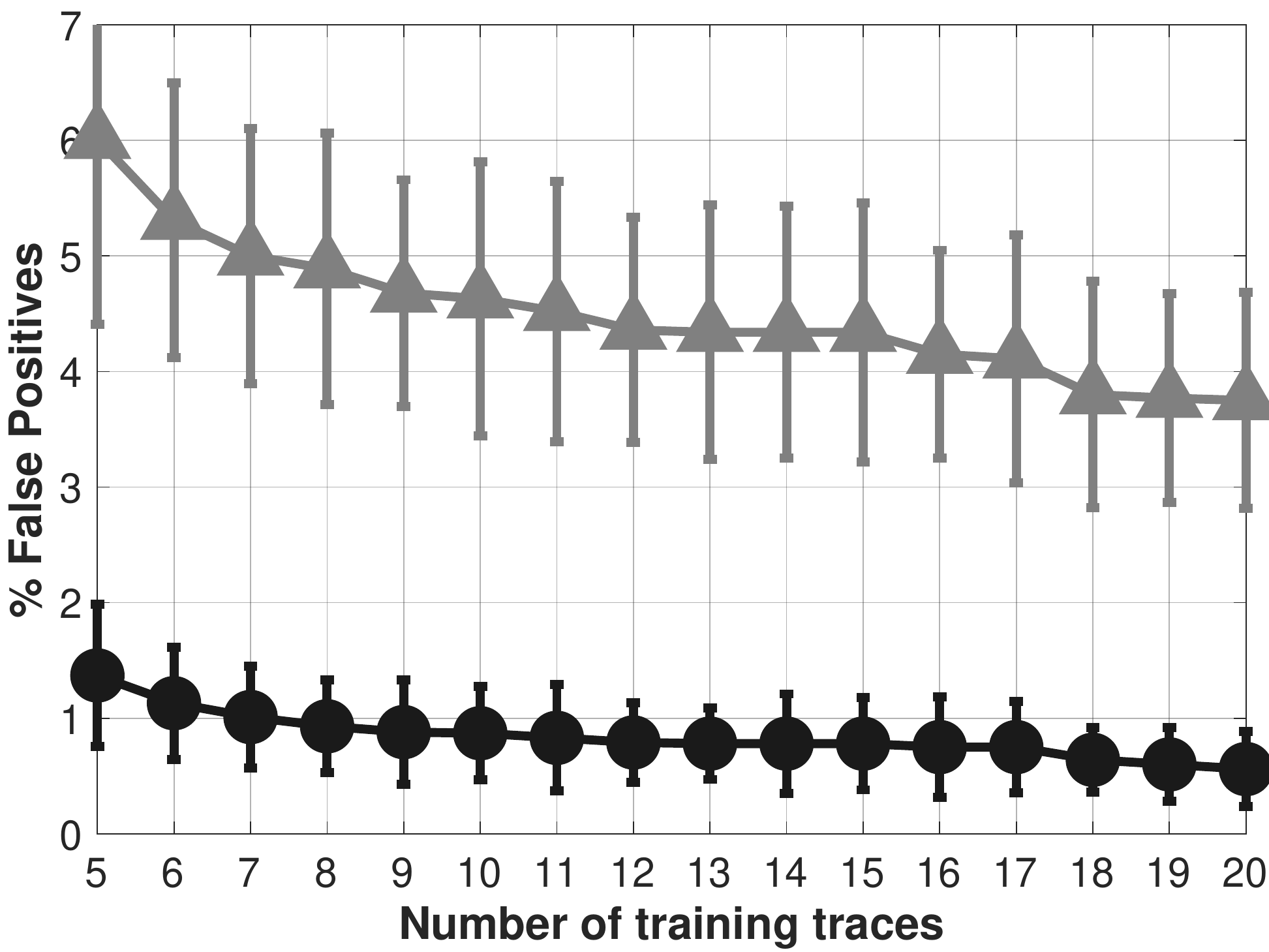}
    }
    \subfloat[Network management workload (NET).\label{fig:fp_network_wl}]{%
        \includegraphics[width=0.67\columnwidth]{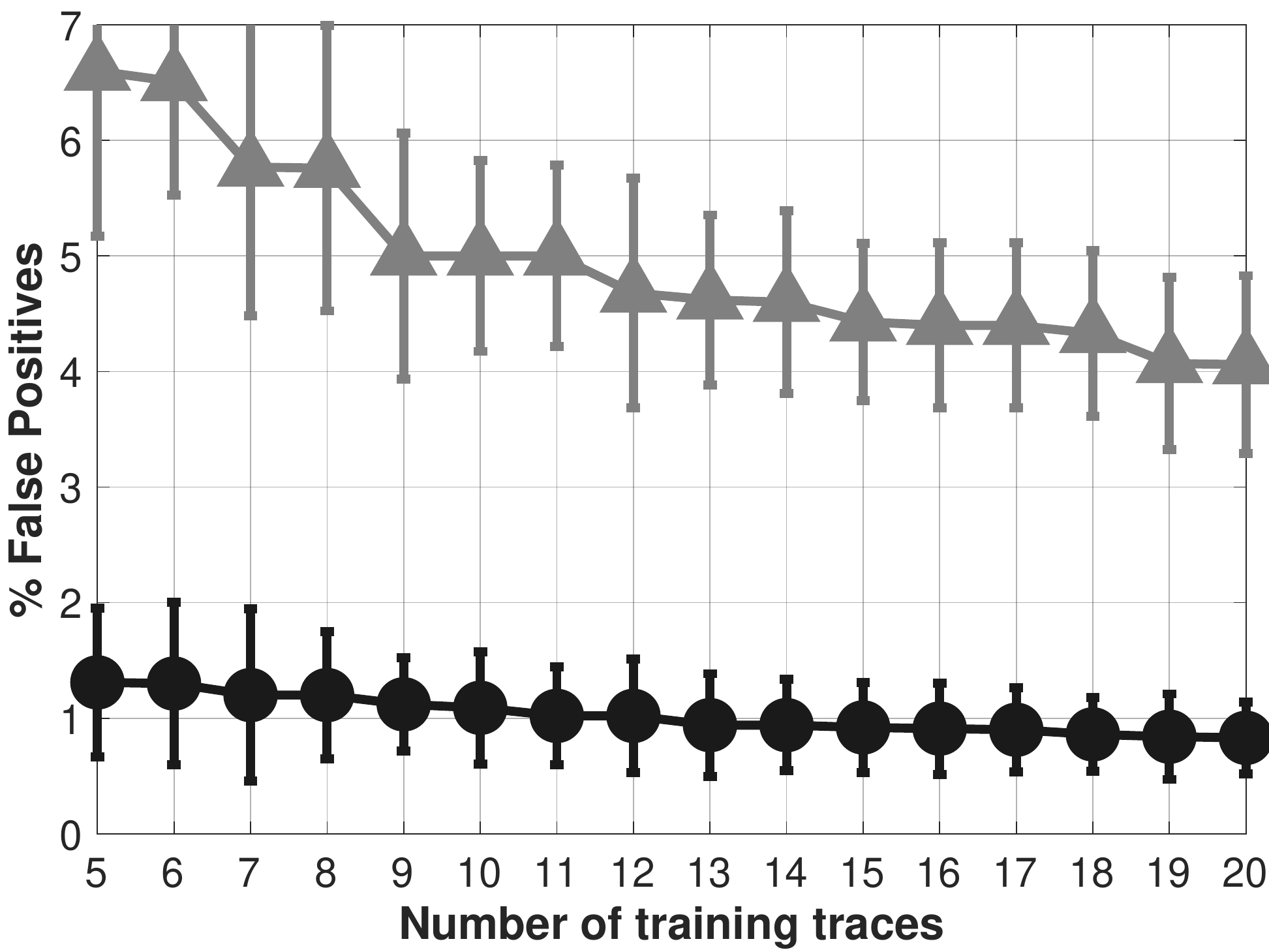}
    }
    \subfloat[Storage management workload (STO).\label{fig:fp_storage_wl}]{%
        \includegraphics[width=0.67\columnwidth]{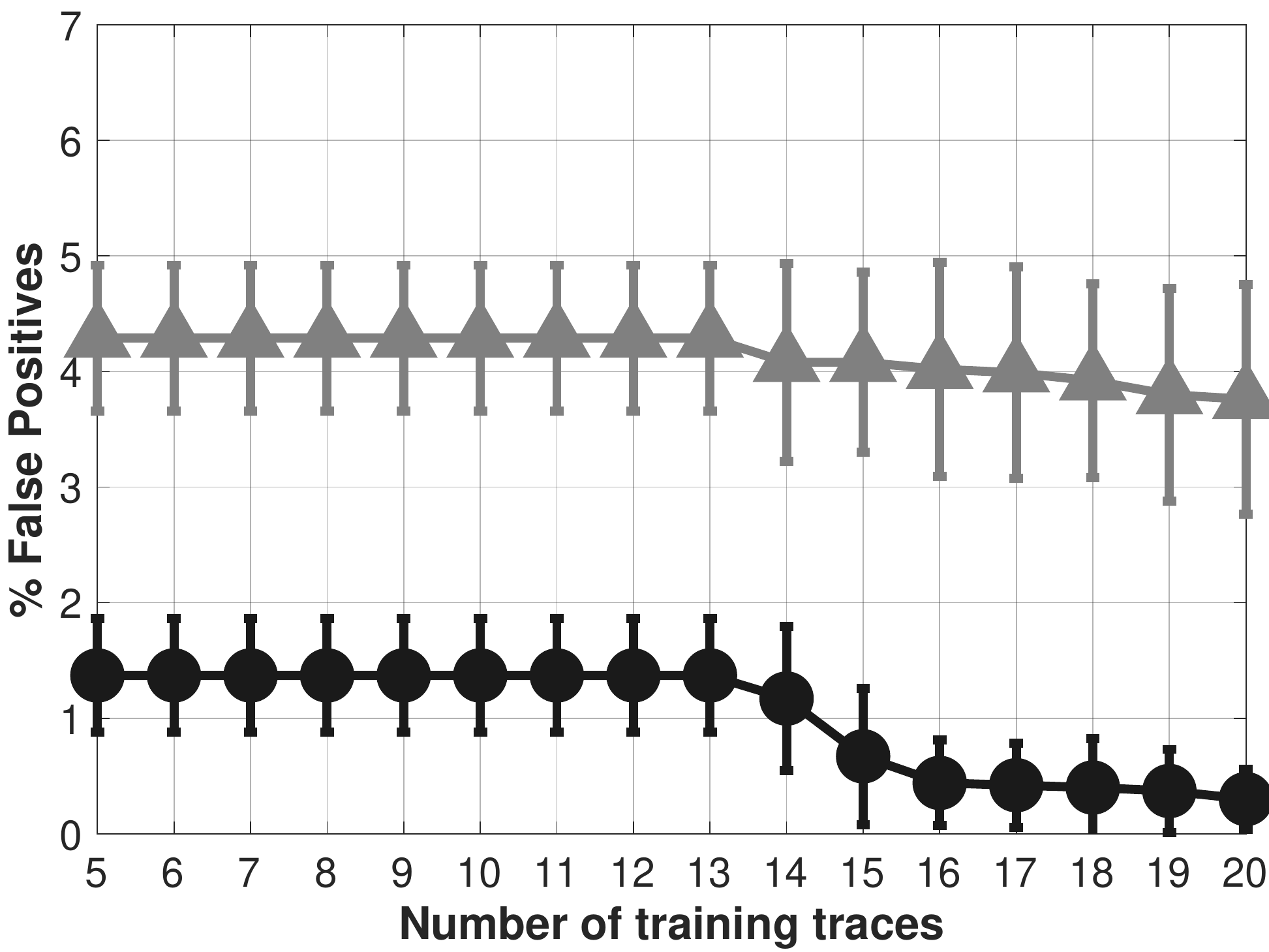}
    }
    \vspace{-0.3cm}
    \caption{False positives analysis.}
    \label{fig:fp_analysis}
    \vspace{-0.6cm}
    
\end{figure*}

We evaluated the accuracy of the proposed approach in terms of \emph{false positives} and \emph{false negatives}. False positives are non-anomalous events that are mistakenly labeled as anomalous (either spurious or missing) by the proposed approach. False negatives are anomalous events that are not identified by the approach (i.e., they are labeled as non-anomalous).

Our experiments generated about half a million events across more than two thousands of execution traces. A key concern for evaluating anomaly detection is the need for a reliable ground truth about the actual label of the events (anomalous or non-anomalous). Unfortunately, manually assigning labels to such a large set of data is prone to errors and unfeasible in practice. 
Thus, we adopt an automated approach. In order to understand which suspicious event (i.e., spurious or missing events) is not actually anomalous, we performed an analysis using an increasing number of sequences of distinct fault-free executions. Since such executions represent the normal behavior of the system, every anomaly identified by the approach should be considered as a false positive.


As a term of comparison, we consider both the full approach, denoted as ``LCS with VMM'', and a baseline approach denoted as ``LCS''. The ``LCS'' approach represents a simplistic approach to failure analysis that just aligns and compares traces without using a probabilistic model to account for non-deterministic variations. In this way, we can separately evaluate the relative impact on the accuracy of LCS and of the VMM.


For each workload, we collect a set of fault-free traces an order of magnitude larger than the set of trace used to train the model. Then, we randomly choose $n$ traces to train the model before evaluating the approach with other $m$ distinct fault-free traces. The $n$ training traces and the $m$ test traces are always disjoint sets. 
We vary the number of training traces (i.e., $n=5, 6, ..., 20$) in order to evaluate the impact of the size of the training set on the accuracy of the approach. Furthermore, we perform tens of repetition for each fixed value of $n$, with different random selections of the training traces and of the test traces. For each repetition, we compute the percentage of anomalies (either spurious or missing events) with respect to the length of the compared sequences. This provides us a metric to evaluate the ratio of the false alarms over the total number of events.

Figure~\ref{fig:fp_analysis} shows, for each workload, how the average percentage of false positives varies with the number $n$ of training traces. For each data point, the sub-figures show a vertical error bar representing the standard deviation of the percentage of false positives across repeated evaluations.
We found that increasing the number $n$ of training traces brings an incremental reduction of the percentage of false positives. In all cases, the percentage settles around 1\% on average for ``LCS with VMM''. The simpler ``LCS'' approach has a higher percentage of false positives, which exceeds the 6\% in the case of the first two workloads, and 4\% in the case of the third workload.

\begin{table}[t]
    \centering
        \caption{False negatives analysis.}
        \label{tab:false_negatives}
        \begin{tabular}{ c c c }
            \toprule
            \textbf{Workload type} & \textbf{LCS} & \textbf{LCS with VMM}\\
            \midrule
            DEPL & 6.35\% & 7.00\% \\ 
            NET & 0\% & 0.91\% \\ 
            STO & 0.88\% & 1.37\% \\
            \bottomrule
        \end{tabular}
        \vspace{-0.6cm}
\end{table}

We can also see that the ``LCS with VMM" approach is less sensitive to the size of the set of training traces when the workload is very extensive (i.e., DEPL and NET).
Indeed, the average percentage of false positives is reduced by at most 1\% with respect to the case of a small training set ($n = 5$), while the ``LCS'' exhibits a wider variation. Our approach provides a percentage of false positives almost stable for all the number of training traces, regardless of the workload; moreover, the uncertainty intervals are overlapping for all values of $n$. The lower sensitivity to the size of the training set makes the ``LCS with VMM'' approach more predictable and easier to apply in practice.


Overall, the curves point out that the VMM can improve accuracy, especially for lower sizes of the training set. The difference of accuracy between the VMM and the plain LCS is wider when the number of events in the workload is higher (DEPL and NET): in fact, in these cases, the LCS string comparison technique generates a high percentage of false positives due to the difficulty at aligning a large number of events with several differences in the middle of the traces. In the case of a workload with a shorter number of events (STO), the LCS analysis provides a lower percentage of false positives and, thus, the gap with the VMM is less pronounced.


To evaluate false negatives, we focus on the experiments that experienced at least a failure. We remark that we consider an experiment as failed if at least one API returns an error or if there is at least an assertion check failure during the execution of the workload.
Since the VMM is applied in pipeline after the LCS, there is a risk that the VMM misclassifies an anomalous event as non-anomalous, thus neglecting the failure-related events (i.e., a false negative). In the ideal case, the percentage of false negatives for the VMM matches the LCS. We expect that, if the VMM is accurate enough, the percentages of false negatives for the VMM and for the LCS should be very close.

Before applying the LCS and VMM approaches, we conservatively remove all the uncertain event types that were marked as false positives in the previous analysis in at least one case. Afterward, if at least one of the remaining events is identified as an anomaly, then we consider such an event as a true anomaly. Otherwise, if there is no true anomaly, we consider the experiment as a case of a false negative. We evaluate the false negatives by measuring the percentage of failures with no anomalies reported over all the experiment that experienced a failure.

\begin{figure*}[ht!]
\centering
\subfloat[Training time while increasing the number of the training traces.\label{fig:training_time_traces}]{%
\includegraphics[width=0.5\columnwidth]{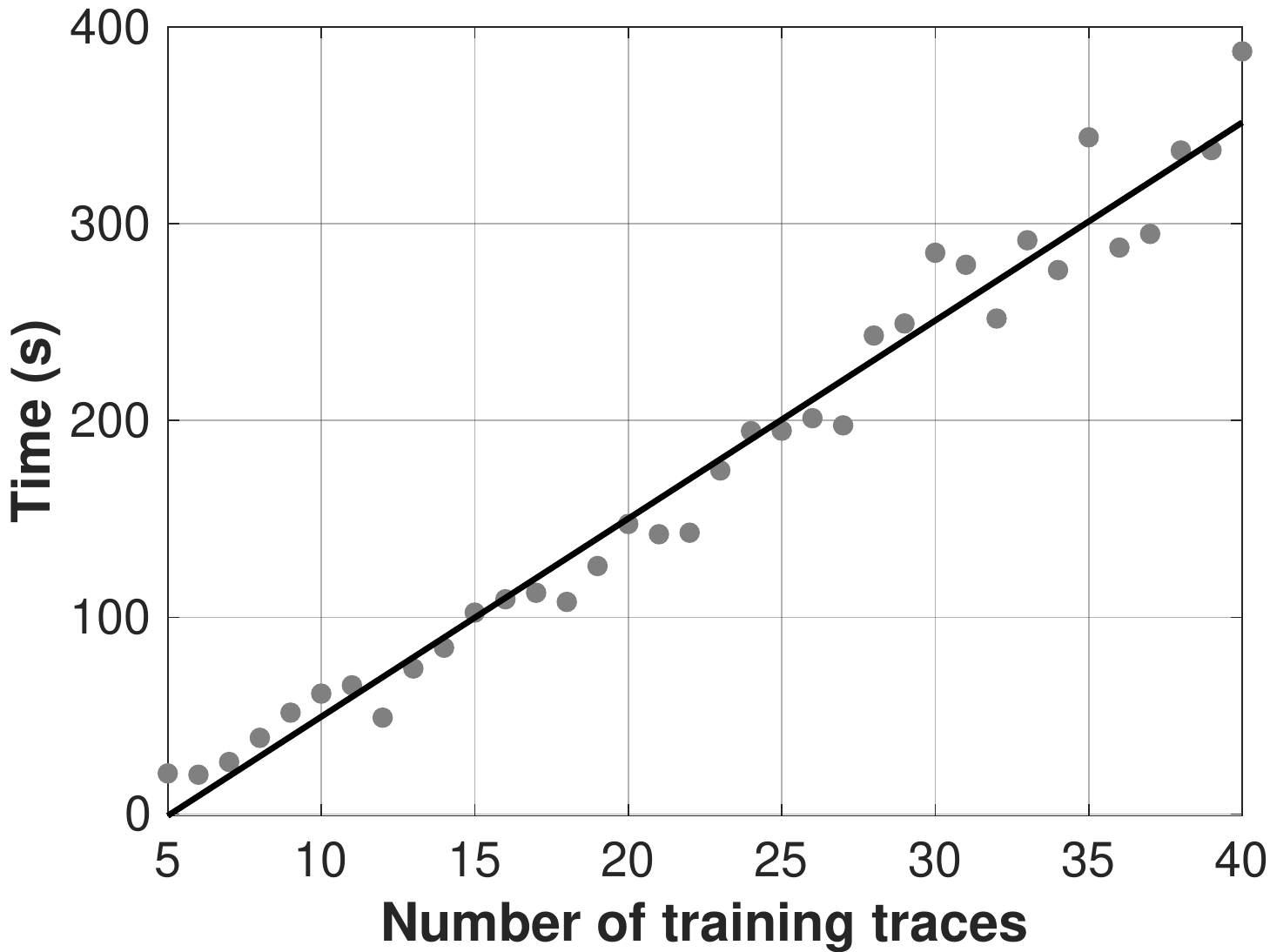}}\hfill
\subfloat[Classification time while increasing the number of the experiments.\label{fig:analysis_time_experiments}]{%
\includegraphics[width=0.5\columnwidth]{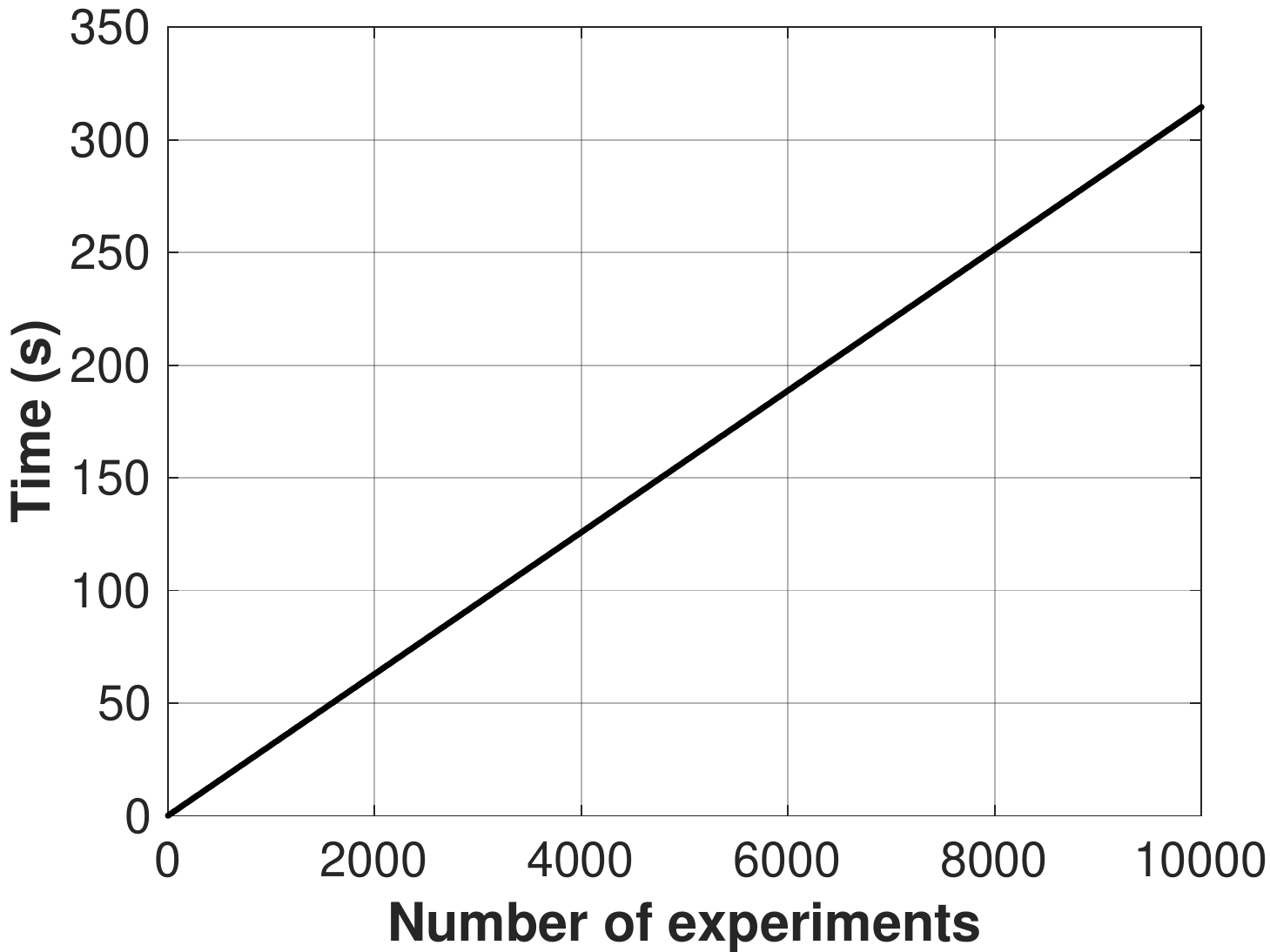}}\hfill
\subfloat[Training time while increasing the size of the traces.\label{fig:training_time_events}]{%
\includegraphics[width=0.5\columnwidth]{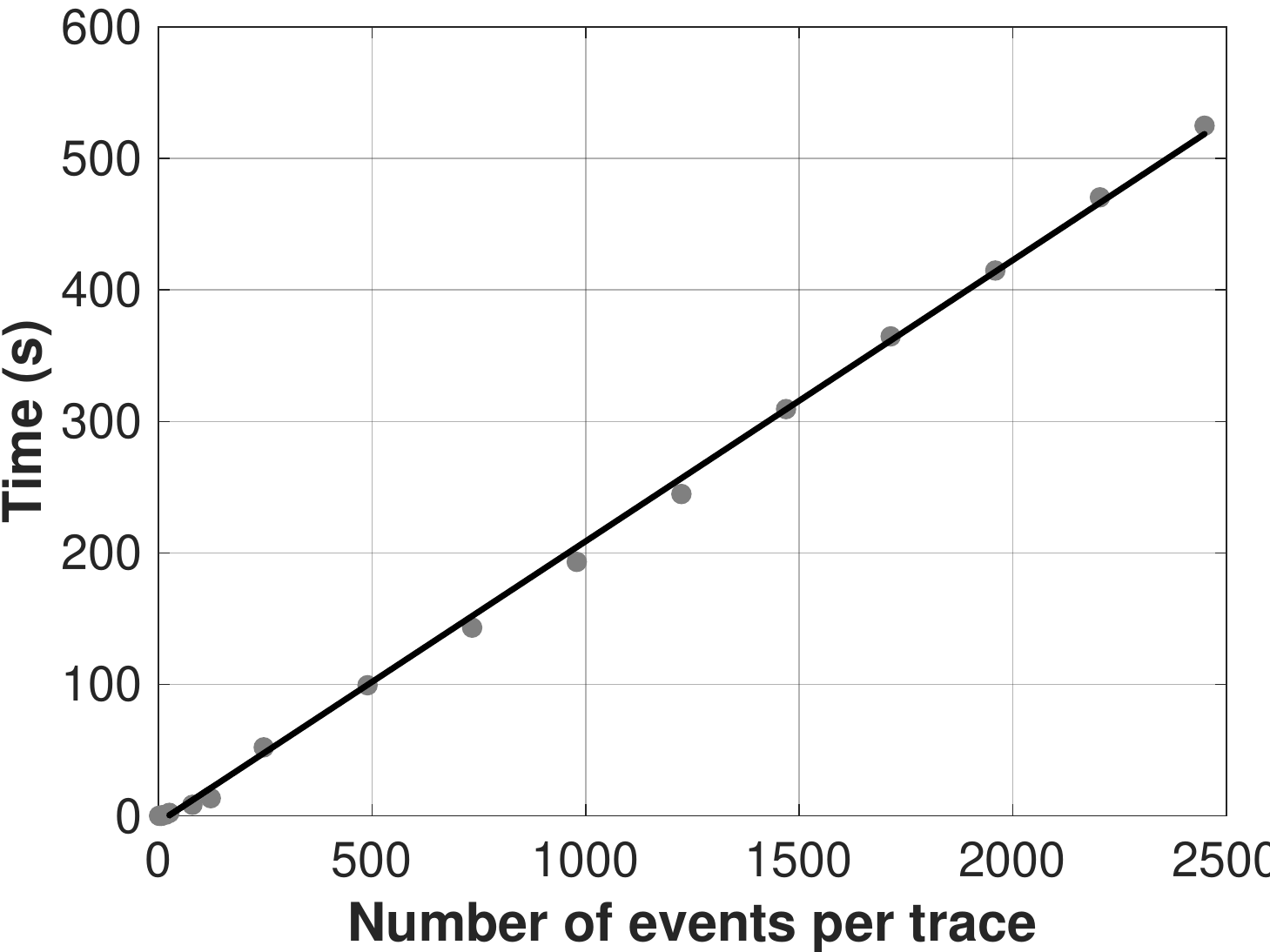}}\hfill
\subfloat[Classification time while increasing the size of the traces.\label{fig:analysis_time_events}]{%
\includegraphics[width=0.5\columnwidth]{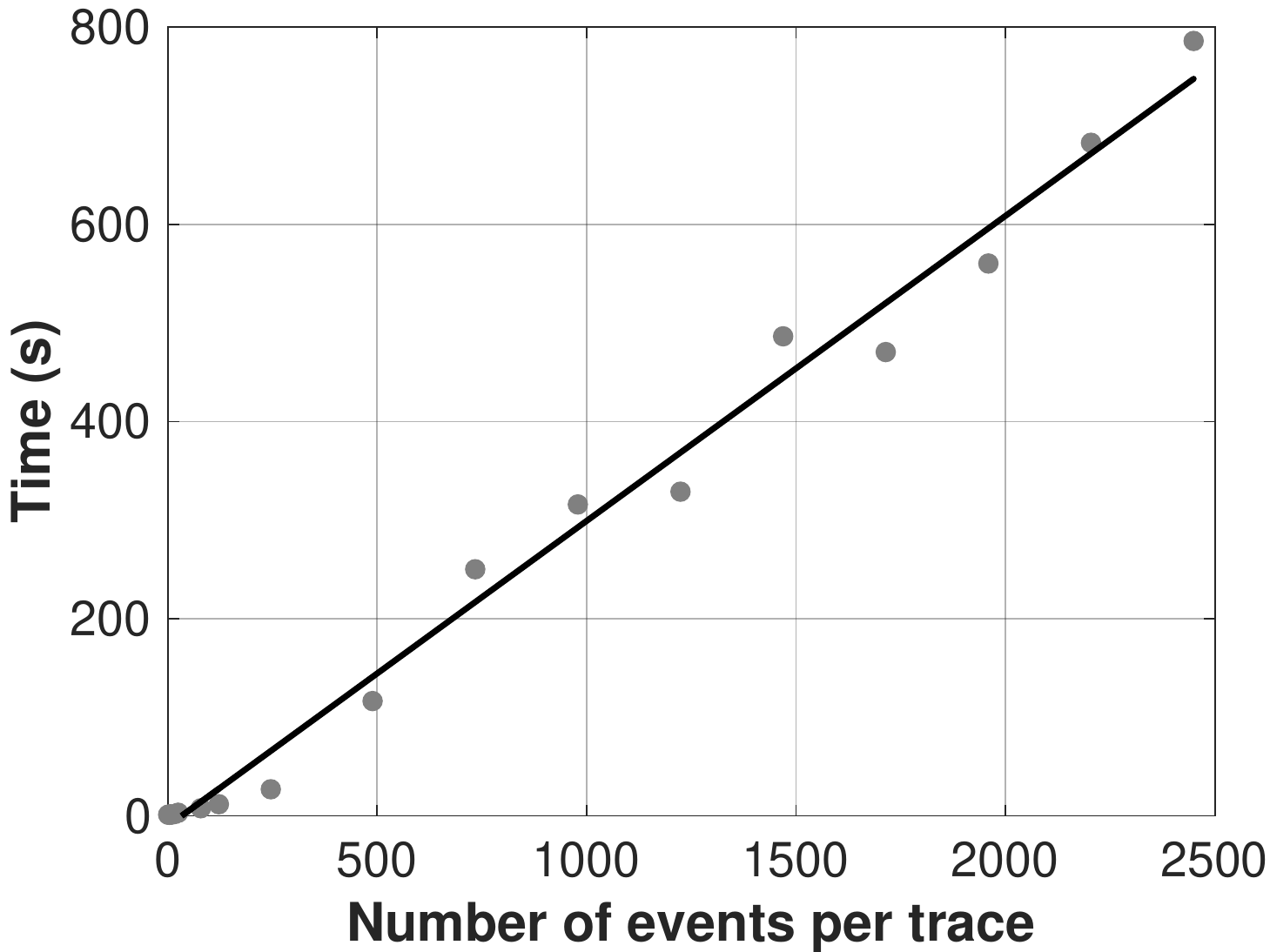}}
\vspace{-0.1cm}
\caption{Computational cost of the proposed approach.}
\label{fig:computational_time}
\vspace{-0.6cm}
\end{figure*}

This method of evaluation is very conservative since we are entirely removing event types that could lead to false positives. Even if these events could have represented true anomalies in some experiments, we ignore anomalies raised for these events, thus restricting the chances of the VMM to point out the failure. However, this approach assures that we do not over-estimate the ability of the VMM at identifying true anomalies since we only take into account anomalies for events that were never affected by false alarms in our previous extensive analysis. 


We executed this analysis for each workload and for different choices of the number $n$ of training traces ($n = 5, 10, 15, 20$). 
Table~\ref{tab:false_negatives} shows the percentage of experiments that experienced a failure, and where the failure was not detected by the approach. This metric is computed by evaluating the number of failed tests in which the algorithm points out no true anomalies for the experiment. The metric is not necessarily $1$ for the ``LCS'' approach since there were failures in our fault injection experiments in which there were neither omitted nor spurious messages (e.g., $6.35\%$ of experiments in DEPL). This behavior happened in the case of ``local'' failures of individual OpenStack components, which did not perform the expected job, but still sent and received the same set of messages of fault-free runs.

We found that the percentage does not vary for different values of $n$ (therefore, the table only reports one value per configuration). Moreover, we found that ``LCS'' and ``VMM'' provide similar results (the differences is always lower than 1\% for all the workloads). It is important to recall that this is an ideal case: since VMM is applied in cascade after LCS, the VMM cannot identify new failures beyond the ``suspect'' events pointed out by LCS. Instead, there is a risk that the VMM filters out some of these events, potentially causing false negatives. However, this is never the case, as the ``VMM'' always raises at least one true positive, even if few anomalies are mistaken as false negatives. This result highlights that the proposed approach can avoid many false positives with a negligible risk of missing a test with a failure. This result is valuable for human analysts since they can focus their debugging activities on a few, specific events that are actually failure-related.


Moreover, we looked in detail at the log messages that were generated by the tests for which we had false negatives. We found that false negatives only occur in failures with no propagation across components. For example, a failure during the cleanup of a resource at the end of the workload typically does not affect any other subsequent operation. For this reason, there is a higher percentage of false negatives in the DEPL workload, since this workload creates (and, thus, deletes) more resources. Conversely, the approach reveals 0.91\% cases of false negatives in the NET workload since it does not contain any cleanup operation, suggesting that the approach is very accurate in terms of false negatives.


\subsection{Performance evaluation}
\label{subsec:performance}

We evaluated the computational cost of the proposed approach, by measuring the time taken for analyzing the event traces, both at training and at classifying them. We performed the analysis with respect to increasing volumes of data, i.e., by varying the number of traces to analyze and the number of the events per-trace (i.e., the length of the sequences of symbols). 

Figure~\ref{fig:training_time_traces} shows how the time to train the model grows with the number of training traces. 
As described in the Subsection~\ref{subsec:accuracy}, when the number of training traces increases, the accuracy of the training model improves. However, a large number of fault-free traces increases both the time for executing more fault-free runs, as well as the computational time for the data analysis, with an approximately linear trend. 
However, the computational cost imposed by VMM seems small enough for practical purposes, since the duration of the data analysis is close to 5 minutes for up to 40 training traces.

Figure~\ref{fig:analysis_time_experiments} shows the time for processing an increasing number of fault injection tests (up to $10,000$ fault injection tests), for a fixed number and size of training traces ($N=10$). The duration increases linearly, but with a lower trend than the previous cases. This is due to the fact that most of the computational cost of the VMM algorithm comes from the training phase, while the estimation phase takes a relatively small amount of time. Therefore, the VMM approach can scale well for high numbers of fault injection tests.

Figure~\ref{fig:training_time_events} gives an indication of the time to train the model when the number of events per trace increases, for a fixed number of training traces ($N=10$). 
The computational cost also grows linearly with the length of the traces, with a slope similar to the previous analyses. In the worst case of thousands of events per trace, the duration of the training can grow up to about 6 minutes, which is still a reasonable duration. 

Finally, Figure~\ref{fig:analysis_time_events} shows the duration of event classification, when the number of events per trace increases, for a fixed number of traces to analyze ($2,000$ traces). In the worst case, the duration of the training can grow up to about 13 minutes.

%% file: related.tex

%

Research studies on debugging distributed systems lead to a variety of \emph{profiling} techniques to pinpoint bugs and performance bottlenecks. 
Aguilera et al. \cite{aguilera2003performance} collect black-box network traces of communications between hosts, in order to analyze requests as they move through the system (e.g., web requests across the tiers of a web application). Their approach infers causal paths of the requests, by tracing call pairs (i.e., request messages, and their corresponding responses), and by analyzing statistical correlations. However, this approach focuses on synchronous (RPC-style) interactions between components, and it is not meant to analyze asynchronous interactions (i.e., the server immediately replies to a request, before issuing causally-related requests and performing more work) and rare events (as the approach focus on the most frequent interactions).

Magpie \cite{barham2003magpie} and Pinpoint \cite{chen2004path} reconstruct causal paths by using more sophisticated tracing infrastructures, by tracing detailed events at the OS-level and at the application server level. The tracing tags incoming requests with a unique \emph{path identifier}, and associates resource usage throughout the system with that identifier. This fine-grain tracing approach does not rely on statistical inference and can provide high accuracy, but it also brings considerable complexity, which makes it difficult to deploy it in practice, especially when considering cloud computing infrastructures with many heterogeneous components (e.g., OSes, middleware, interpreters, etc.).

Gu at al. \cite{gu2018kerep} proposes a methodology to extract knowledge on distributed system behavior of request processing without source code or prior knowledge. The authors construct the distributed system's component architecture in request processing and discover the heartbeat mechanisms of target distributed systems.

Pip \cite{reynolds2006pip} is a system for automatically checking the behavior of a distributed system against programmer-written expectations about the system. Pip provides a domain-specific expectations language for writing declarative descriptions of the expected behavior of large distributed systems and relies on user-written annotations of the source code of the system to gather events and to propagate path identifiers across chains of requests. This approach provides flexibility for the analysis but requires access to the source code, and non-negligible efforts to annotate it.

More recent studies contributed to tools resembling debuggers, but for distributed systems. Pensieve \cite{zhang2017pensieve} is an approach for producing the path to failure, in a similar way to delta debugging: it combines static analysis, and re-execution of the system with iteratively-refined logging, in order reconstruct the intermediate path backward from the failure to the user inputs and events that cause the failure. 
Friday \cite{geels2007friday} is a distributed debugger that allows developers to replay a failed execution of a distributed system, and to inspect the execution through breakpoints, watchpoints, single-stepping, etc., at the global-state level. 
ShizViz \cite{beschastnikh2016debugging} is an interactive tool for visualizing execution traces of distributed systems, which allows developers to intuitively explore the traces and to perform searches; moreover, the tool provides support for comparing distributed executions with a pairwise comparison, even if without probabilistic techniques to filter-out benign variations due to non-determinism.

Recent fault injection solution addressed cloud computing systems. The \emph{Fate} \cite{Gunawi2011a} tool, and its successor \emph{PreFail} \cite{Joshi2011b}, simulate disk failures, network partitions, and crashes of nodes, by exploring multiple occurrences of faults during the same experiment, to test recovery procedures more thoroughly (e.g., at tolerating further network/disk faults occurring during recovery). To address the combinatorial explosion of experiments, these tools adopt user-programmable policies to prune redundant experiments (e.g., injections in symmetric states or in paths that were already covered). 
Ju et al. \cite{Ju2013a}, \emph{ChaosMonkey} \cite{chaos_monkey}, and \emph{Jepsen} \cite{jepsen} test the resilience of cloud infrastructures by injecting crashes (e.g., by killing VMs or service processes), network partitions (by disabling communication between two subnets), and network traffic latency and losses. 
\emph{CloudVal} \cite{Pham2011} and Cerveira et al. \cite{cerveira2015recovery} use fault injection (CPU and memory corruptions, resource leaks) to test the isolation among hypervisors and VMs. Pham et al. \cite{pham2017failure} applied fault injection on OpenStack to create signatures of the failures, in order to support problem diagnosis when the same failures happen in production. 
Once fault injection reveals a failure, in most cases it is the tester's responsibility to look at what happened during the test, and come up with an interpretation of the issue and of a potential solution to make the system more fault-tolerant.

Our approach differs from anomaly detection solutions using ML models or employing self-adapted monitoring \cite{alonso2011predicting,sauvanaud2016anomaly,ehlers2011self}, and it is unique in the design space of distributed debugging tools. To the best of our knowledge, this is the first approach that applies distributed debugging techniques for interpreting the fault injection experiments. In the context of fault injection, the fault-free executions are used as a reference for identifying anomalies in fault-injected executions performed under the same conditions (same workload, same node deployment, etc.): therefore, the approach does not rely on programmer-written specifications to identify failures (even if such specifications could cooperate with our approach to gain further insights); moreover, our approach does not rely on inferring causal relationships (which requires more intrusive instrumentation and may be inaccurate for asynchronous and rare interactions). Since the approach only relies on modeling the observed sequences of events, it can be easily deployed and integrated into interactive tools for debugging and visualization, to provide more robust trace comparison and analysis abilities.

%% file: conclusion.tex

In this paper, we propose a technique for analyzing execution traces of distributed systems under fault injection, by comparing the executions to fault-free ones in order to point out anomalies. To address the problem of non-determinism (which may lead to ``benign'' anomalies not actually related to failures) we develop a sequence comparison approach supported by a probabilistic model. The probabilistic model is built from a group of several fault-free execution traces, in order to reflect ``benign'' variations that normally occur in the distributed system. Moreover, to make the approach applicable to black-box systems and not reliant on intrusive instrumentation, we base our probabilistic model only on externally-observable traces of messages, which are analyzed as sequences of symbols using Variable-order Markov Models. We evaluated the approach within the OpenStack cloud computing platform: we found that the VMM limits the false positives compared to a non-probabilistic comparison of execution sequences, without significant loss in terms of false negatives. Moreover, the VMM is lightweight enough to be applicable with a low computational cost. Future development is to integrate the approach with tools for fault injection and debugging, such as for reporting the anomalies to the users and for clustering fault injection tests to better support the human analysts.